# Software Implementation Level Countermeasures against the Cache Timing Attack on Advanced Encryption Standard

Udyani Herath, Janaka Alawatugoda, and Roshan Ragel, *Member, IEEE*

*Abstract-* **Advanced Encryption Standard (AES) is a symmetric key encryption algorithm which is extensively used in secure electronic data transmission. When introduced, although it was tested and declared as secure, in 2005, a researcher named Bernstein claimed that it is vulnerable to side channel attacks. The cache-based timing attack is the type of side channel attack demonstrated by Bernstein, which uses the timing variation in cache hits and misses. This kind of attacks can be prevented by masking the actual timing information from the attacker. Such masking can be performed by altering the original AES software implementation while preserving its semantics. This paper presents possible software implementation level countermeasures against Bernstein's cache timing attack. Two simple software based countermeasures based on the concept of "constant-encryption-time" were demonstrated against the remote cache timing attack with positive outcomes, in which we establish a secured environment for the AES encryption.**

## I. Introduction

In the modern world, symmetric key cryptography possesses a high esteem where it provides secure communication in the presence of unauthorized third parties. In computerized information systems, symmetric key cryptosystems are used to provide confidentiality for sensitive messages by encrypting them into messages that make no sense to third parties. Symmetric key cryptosystems are equipped with two public algorithms known as the encryption algorithm and the decryption algorithm. Encrypting is an invertible mathematical operation performed on the sensitive message with the aid of a public encryption algorithm and a secret-key which is shared amongst authorized parties.

When the encrypted message comes to an authorized person who has the knowledge of the secret-key, she/he can invert the encrypted message with the aid of a public decryption algorithm and the secret-key to get the meaningful sensitive message.

Udyani Herath is with the Department of Statistics and Computer Science, Faculty of Science, University of Peradeniya, Peradeniya 20400, Sri Lanka (email: udyani.herath@gmail.com)

Janaka Alawatugoda is with School of Electrical Engineering and Computer Science, Queensland University of Technology, Brisbane, Queensland 4000, Australia (email: araliyaqut@gmail.com)

Roshan Ragel is with the Department of Computer Engineering, Faculty of Engineering, University of Peradeniya, Peradeniya 20400, Sri Lanka (email: ragelrg@gmail.com)

Semantically, it is assumed that an unauthorized person cannot learn any useful information about the sensitive message by seeing the encrypted message even though both the encryption and the decryption algorithms are public.

The Advanced Encryption Standard (AES) is such a symmetric key cryptosystem, which has been adopted by United States National Institute of Standards and Technology (NIST). This was developed by Belgian cryptographers, John Daemon and Vincent Rijmen and originally called Rijndael; and was chosen as a replacement for the Data Encryption Standard (DES) after finding it was less secure [1].

AES is established on a design principle recognized as a substitution permutation network, and is fast for implementation both in hardware and software.

AES is a symmetric block cipher with a block size of 128 bits (16 bytes) and comprises one of the three key lengths of 128, 192, or 256 bits, which use 10, 12 and 14 rounds respectively. Each round except the last, involves four stages: Sub Bytes, Shift Rows, Mix Columns and Add round key. AES formats plaintext in to 16-byte blocks, and then treats each block as a 4x4 state array. For each operation, AES implementation uses pre-computed tables of values. Therefore, time consumed for each operation depends on those table lookups; hence the vulnerability to side channel attacks is increased.

A Side Channel Attack (SCA) is any attack based on "side channel information", the information which can be gained from encryption device, which we cannot consider as the plaintext to be encrypted or the cipher text that results from the encryption procedure. It is not an attack based on theoretical weakness of the algorithm but of the implementation. Attackers use information leaking from the system such as timing information, electromagnetic beats and sometimes it can be even a sound [2].

Using cache timing information as the side channel, cache timing attacks are performed by analysing leaking timing information when a crypto system is performing encryption. Cache, which is a component that is used by the central processing unit of a computer, reduces the average time to access memory and store data providing a faster access for future needs. The data stored in the cache can be the value accessed recently or the duplicate of original values that are stored somewhere else in the computer [3].

When a particular data is requested, if it is stored in the cache, this request can be assisted by merely reading the cache (aka. cache hit) and it will be much faster compared to the data to be recomputed or drawn from its original storage location (aka. cache miss) [4].

In AES implementation, pre-computed tables or the S-boxes are stored as arrays and since the cache memory has limited space, they cannot be fully loaded in to the cache memory. The program accesses these array entries while the encryption is done, and in these instances, cache hits or misses can occur, leading to a significant timing variation, hence leaking timing information for any attackers trying to break the system.

In 2005, Bernstein revealed a remote cache timing attack against AES and showed that AES is susceptible for timing attacks [5]. In our study, we follow his proposed attack as our guide, where we have analysed and implemented it in our environment and we have created some countermeasures and have assessed their performance and reliability.

The rest of the paper is organized as follows. Section II presents related work. In Section III, we have described our implementation of Bernstein's attack. Section IV is on countermeasures against Bernstein's attack where we have tested two implementation level countermeasures by enforcing constant number of clock cycles for encryption of a data packet. We could practically demonstrate the success of proposed countermeasures by showing the attacker's inability to find the actual secret-key. Further, we have appraised the efficiency of our suggested countermeasures comparatively with the efficiency of unmasked AES implementation. In section V, we have included the performance influence of our explored countermeasures. In Section VI we have concluded the paper.

## II. RELATED WORK

Bernstein performed his attack successfully by using the OpenSSL 0.9.7a AES implementation on an 850MHz Pentium III Desktop Computer, running FreeBSD-4.8 as a network server. He extracted the complete AES key using a client machine and pointed out that the same technique can be performed on more complicated servers with additional timing information. He has also tested an AMD Athlon, an Intel Pentium III, an IBM PowerPC RS64 IV and a Sun UltraSPARC III processor with positive results.

In the meantime, Kocher has performed timing attacks on implementations of Diffie-Hellman, RSA, DSS and other crypto systems [6]. He stated that timing attacks are centred on measuring the time it takes for a unit to perform operations, where it lead to information about secret keys and break the crypto system. He also stated that the attack is computationally not much difficult and most of the time only known cipher text is required and he has presented some techniques for preventing the attacks.

Felten et al. [7] have done timing attacks on web privacy. They have described a class of attacks that can compromise the privacy of user's web-browsing histories. The attacks permit a malicious web site to collect information on users' browsing activities. By assessing the time the user's browser requires to perform certain operations it can be determined. According to them, the time required for operations depends on the user's browsing history and this time variations bear enough information to contain user's privacy because of the various forms of caching performed by browsers. According to them these attacks can be carried out without the victim's knowledge. They pointed out that simple countermeasures cannot prevent these types of attacks. Therefore they have proposed a way of re-creating browsers.

In 2011, Alawatugoda et al. [8] have performed a research on countermeasures against remote cache timing attacks by planning and implementing few possible countermeasures. In order to prevent cache timing attacks, they have followed the approach of masking leaking timing information. They have added several code fragments in to the AES implementation. They have been able to do it without changing its semantics and also without severely reducing the competence of it. The software based countermeasures they have tested involve adding randomness and few actions on T-tables such as pre-fetching table values and cache partitioning where cache locations are allocated to load T-tables. They have been able to achieve the target successfully.

In 2012, Jayasinghe et al. have presented constant time encryption as a countermeasure against remote cache timing attacks [9]. Most of the software based countermeasures are vulnerable to statistical analysis though they are flexible and easily organized. With that problem in mind, they have tested a countermeasure that is safe against statistical analysis. Their method reschedules the instructions of AES algorithm where the encryption rounds will consume constant time regardless of the cache hits and misses. They have done so in major three steps which are decomposing the code into smaller bitwise operations, adding each and every bitwise instruction sets to queues and processing each queue. They have shown that the countermeasures have eliminated the side channel vulnerability.

In this paper, we have focused on masking timing information by constant time implementations where AES encryption program is rescheduled to take a constant time execution regardless of cache hits and misses. Bernstein has proposed several countermeasures which are related to the constant time approach and also the practical problems of them. According to Paul Kocher, for masking timing information constant time approach is an approved working method and he focused on Diffie-Hellman, RSA and DSS systems. Alawatugoda et al. has proposed few different methods other than constant time encryption as countermeasures e.g. random or specified '*for*' loops, pre-fetching T-table values and cache partitioning. Jayasinghe et al. has proposed to use pipeline depth of the targeted processor to make timing details constant. We have tried a different, much simpler approach than those mentioned above by considering number of clock cycles and it proved to be successful as well as efficient against cache-timing attacks. In this paper, we are presenting the performance, soundness and effectiveness of our work.

## III. BERNSTEIN'S CACHE TIMING ATTACK

Our initial cryptographic system was established according to Bernstein's work [5]. He has successfully extracted the complete AES key from a network. Two identical servers are used where one is the actual victim's server and the other is a duplicate identical to the victim's server. The server program is compiled on a 733MHz Pentium III Desktop Computer running FreeBSD and the Client is a 2.0 GHz Intel Core i3 laptop machine running Ubuntu 10.10. Client sends random data packets to the server for encryption and the time for encrypting the packets are recorded. Server encrypt those data packets using a private key and sends back the information of encryption time with scrambled zeros to Client. Server avoids sending any encrypted data to the client which acts as the attacker.

The attack has three stages. Initially client collects data under a known key. Secondly it collects data for the unknown key. Both data are saved in separate files. As the final step the key is deduced by correlating the two obtained files and it will produce the possible key space according to the timing details.

Fig. 1 shows the sample key possibilities obtained after correlating is done. The illustrated key space is approximately $3.9 \times 10^{10}$ as opposed to $3.4 \times 10^{38}$ which is the original key space of 128 bits AES.

In Fig. 1 the number of possible keys for each byte and the byte number are shown in first and second columns respectively and the possible keys are shown by the rest. AES key bytes are highlighted.

A brute-force search is done in order to find the AES key from the extracted possible keys. Each possible key is checked against the server's scrambled zeros by encrypting 0s with it and checking whether it is equal with the scrambled zeros. This is done until the search program finds the full AES key.

Following an idea from Alawatugoda's paper [8], we experimented on search time for various sizes of key combinations so that we could derive some important conclusions when the reduced key combinations are attained from timing attacks. The experiments were done using a 2.0GHz Intel Core i3 processor which also acted as our 'Client' in the attack tested. Results are reported in Table 1. And the Fig. 2 shows the graph drawn from the values of Table 1.

TABLE I
SEARCH TIME FOR DIFFERENT KEY POSSIBILITIES

| Possible AES key space | Time taken (s) |
|---|---|
| $10^1$ | $2 \times 10^{-3}$ |
| $10^2$ | $3 \times 10^{-3}$ |
| $10^3$ | $4 \times 10^{-3}$ |
| $10^4$ | $1.2 \times 10^{-2}$ |
| $10^5$ | $4.3 \times 10^{-2}$ |
| $10^6$ | $5.1 \times 10^{-2}$ |
| $10^7$ | $2.95 \times 10^{-1}$ |
| $10^8$ | 2.82 |
| $10^9$ | 28.10 |
| $10^{10}$ | 281.49 ~ 5 minutes |
| $10^{11}$ | 2804.75 ~ 47 minutes |
| $10^{12}$ | 18600.58 ~ 5 hours |

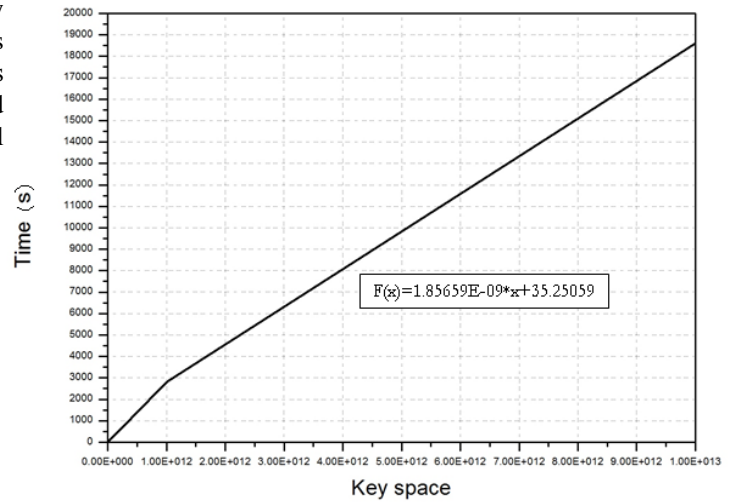

Fig. 2. Graph of key space vs. Search time in seconds

## IV. COUNTERMEASURES

In the literature, there have been many countermeasures proposed and tested. Mainly these can be divided as software based and hardware based countermeasures. In this research, we have focused on a software approach. In order to protect implementations of secret-key cryptographic primitives there are only limited choice of countermeasures. Those are to either masking of information or randomization of operations. We concentrated on masking timing information where it is tried to achieve by constant time encryption and tested two approaches. Detailed explanations of the methods are given below.

| 1 | 0 | f3 |
| 1 | 1 | a9 |
| 8 | 2 | 6c 68 6a 6f 6b 6d 6e 69 |
| 1 | 3 | 51 |
| 17 | 4 | 83 82 8b 8a 32 3a 2b 9b 3b 2a 9a 33 22 93 92 23 b2 |
| 8 | 5 | 89 88 8b 8a 8e 8d 8f 8c |
| 8 | 6 | 0d 0f 0a 0b 0e 09 0c 08 |
| 8 | 7 | 63 60 66 61 62 67 64 65 |
| 1 | 8 | b2 |
| 4 | 9 | e2 e3 e1 e0 |
| 139 | 10 | 7f 78 7e 79 7a 7c 7b 7d ea 6f ed 6a 6d 68 6c ef eb ec e8 6b e9 69 ee 6e c7 c2 c1 c0 c5 c6 c3 c4 a7 a4 4c 49 4e 4f a0 48 a6 a5 4d a2 a3 4b 4a a1 f4 f6 91 f5 f2 f1 f7 90 f3 95 96 94 97 92 93 f0 62 28 2f 2a 5c 61 64 5b 2b 2c 2e 59 67 2d 58 5f 5d 29 65 66 5e 60 ba 63 5a bd be bb bf b9 b8 bc 52 50 45 47 57 44 54 42 40 51 46 56 82 55 e1 53 85 e0 43 |

Fig. 1. Sample key spaces after correlation

*A. Fixed number of clock cycles*

The AES software implementation is rescheduled such that it will take constant time for execution. Algorithm 1 shows the part of the code executed for the first round inside *AES_encrypt* function in AES implementation. Initially a fixed number is defined (line 1). By using a time stamp, number of clock cycles for each round is obtained (lines 2-7). Then the difference between the two is obtained (line 8). That is, the fixed number and the number of clock cycles for each round. Then, a loop is included where it runs from zero to the number that is obtained (line 9). This is done for each and every 16 rounds.

```
1. static int fixed_cycles= 250;  //a fixed number of clock cycles
2. int j=0;
3. int startT[THE_SIZE];  //the start of the timestamp
4. static int num_cycles[THE_SIZE]; //array for number of clock cycles
5. int cyc_diff[THE_SIZE]

/* round 1: */

6. startT[0] = timestamp1();

……round 1 AES implementations……

7. num_cycles[0] = timestamp1() - startT[0];
8. cyc_diff[0] = fixed_cycles - num_cycles[0];

9. for(j=0; j< cyc_diff[0]; j++)
                       {
                                   asm("nop");
                       }
```

Algorithm 1. Fixed number method

Fixed number method has taken around 5630 additional clock cycles than the usual. That is, it takes around 5050 clock cycles for original AES implementation when calculating the average number of cycles for encryption of 800 byte packets whereas this changed AES execution consume around 10700 clock cycles. We observed that it is approximately 2.1 times higher than the number of clock cycles in the unprotected AES implementation.

In addition, it was detected that some of the key bytes are missing after the usage of this modified AES implementation. The attacker will be unable to recognize the authentic timing pattern properly because of the incorrect timing information. And also it was observed that the total number of possible keys is nearly $9.4 \times 10^{35}$. That is a considerably larger value when comparing with the original AES implementation which contained around $3.9 \times 10^{10}$ possible keys when attacked. Fig. 3 shows the results marked in the graph of key space vs. search time in seconds. Time estimated from the graph is $5.5 \times 10^{19}$ years. That is, with the attack program tested in this research it will be impossible to find the correct key within reasonable amount of time.

Above mentioned results are obtained when the fixed number was set to 250. It is possible to create countermeasures with different levels of protection by increasing or decreasing the fixed number. As a test search the fixed number was increased to 2500 and it was observed that altered AES implementation costs about 87220 clock cycles. It is a very large overhead compared with the unprotected implementation and it was 17.27 times higher in number of clock cycles. And the possible number of keys obtained was $1.1 \times 10^{37}$.

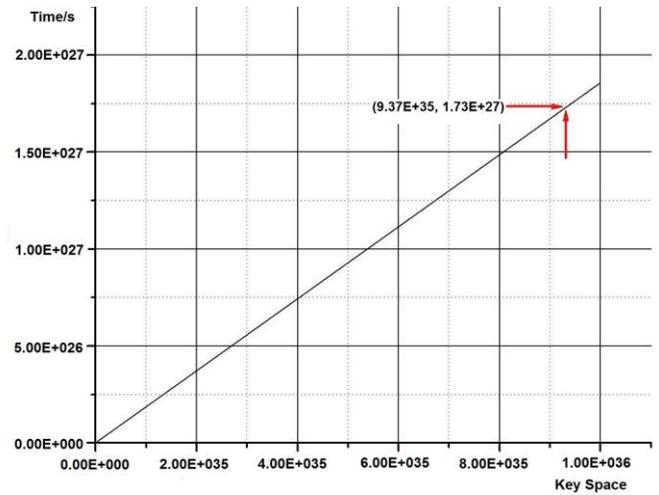

Fig. 3. Result of the fixed number method

*B. Average number of clock cycles*

Instead of defining a fixed number of clock cycles by an outsider (the programmer that is), numbers of clock cycles for rounds themselves are used to calculate an average and to perform constant encryption time by equalling clock cycles up to the averaged value. Algorithm 2 shows the code fragment implemented for the fifth round inside *AES_encrypt* function in AES implementation.

Initially by defining a time stamp, number of clock cycles for each round is obtained (lines 1-7). Then the average is calculated incrementally for each round (lines 8, 9 and 10). In between each round a 'for' loop is included where it execute from zero to the number that is obtained as the difference of averaged value and the clock cycle value of the particular round (line 11).

Since there are 16 rounds in AES algorithm, code fragment in Algorithm 2 runs 16 times, incrementally calculating average for each round when encrypting data packet.

The operation of making the encryption time constant using an averaged value of clock cycles has taken around 1000 more clock cycles than the usual. That is it takes around 5050 clock cycles for original AES implementation when determining the regular number of cycles for encryption of 800 byte packets whereas this altered AES costs about 6050 clock cycles. We observed that it is approximately 1.19 times greater than the number of clock cycles in the unprotected AES implementation.

```
1. int startT[THE_SIZE];  //the start of the timestamp
2. static int num_cycles[THE_SIZE]; //array for number of clock cycles
3. static int Cycle_avg[THE_SIZE]; //the average of clock cycles
4. int cyc_diff[THE_SIZE]; //the difference between average and a clock cycle
5.int sum = 0;

/* round 5: */

6. startT[4] = timestamp1();

……..round 5 AES implementations……..

7. num_cycles[4] = timestamp1() - startT[4];
8. sum = sum + num_cycles[4];
9. Cycle_avg[4] = sum/5;
10. cyc_diff[4] = Cycle_avg[4] - num_cycles[4];

11.for(j=0; j< cyc_diff[4]; j++)
       {
       asm("nop");
       }
```

Algorithm 2. Average number method

After using our AES implementation in the server we noticed that some of the key bytes have been missing. The attack has been unable to recognize the authentic timing pattern properly because of the wrong timing information. The total number of possible keys is nearly $2.1 \times 10^{38}$. That is a considerably larger value when comparing with the original AES implementation which contained around $3.9 \times 10^{10}$ possible keys. Fig. 4 shows the result marked in the graph key space vs. search time. According to the graph the estimated time to find the key is $1.2 \times 10^{22}$ years. That is, it will be impossible to find the correct key even with the attack tested.

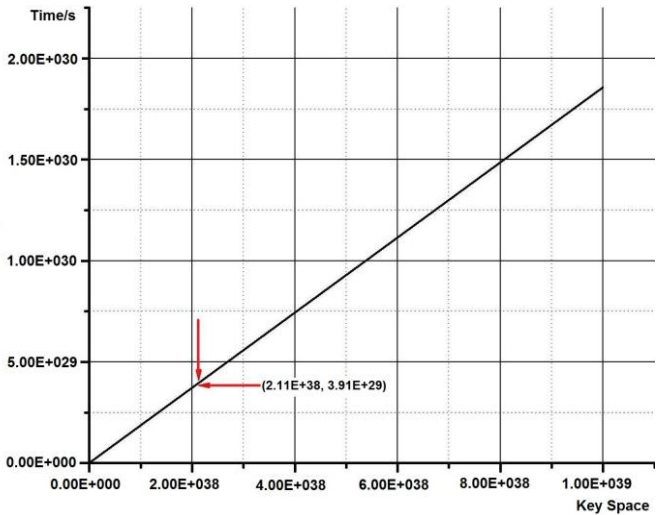

Fig. 4. Result of the average number method

V. PERFORMANCE EVALUATION

Missing a key byte can cause difficulties for an attacker to find the secret key, because even if a key byte is missed and the position of it is known to the attacker, he will have to execute a search on 256 times larger key space. Thus it is hard to perform it. Further, it becomes much harder if the invader does not know the location of the missing key byte. When few key bytes are missed the key space will be greater in some power (number of bytes missing) of 256. Hence if a countermeasure can slip even one key byte the protection of the AES implementation increases considerably.

In this research, the intention was to create simple and effective countermeasures based on the concept of constant encryption time. To achieve that target two different approaches were tested. The two countermeasures have shown varying amounts of overheads. It might directly affect the competence of the AES implementation. To decide which one is better, we can evaluate outcomes of missing key byte or the number of clock cycles compared to the original unprotected AES implementation.

Table II shows the performance evidence of each countermeasure

TABLE II
PERFORMANCE OF TESTED COUNTERMEASURES

| Countermeasure | No. of missing key bytes (m) | Avg. no. of clock cycles per encryption | No. of multiples from the original AES implementation (s)( original val. = 5050) |
|---|---|---|---|
| Fixed number | 3 | 10680 | 2.10 |
| Average number | 2 | 6050 | 1.19 |

Using the information given in Table II, Equation (1) is proposed in [8] to calculate competence of the countermeasures tried. We can get an assessment about the effectiveness of the countermeasures. In the equation (1), *m* indicates the number of missing values and the multiple value comparing with the original AES implementation is given by *s* (or is X times higher than the unprotected AES).

$$Efficiency = \frac{1}{s}m \qquad (1)$$

When the efficiency values of tested countermeasures were calculated using the equation (1), the following results were obtained.

Efficiency (Fixed number) = 1.42
Efficiency (Average number) = 1.68

VI. CONCLUSION

In this research two simple countermeasures based on the concept of constant encryption time were planned and executed against the remote cache timing attack which was suggested by Daniel Bernstein. The efficiencies were calculated using equation (1) by comparing the countermeasures.

According to the research the most effective and efficient countermeasure amongst proposed two countermeasures is using an average number of clock cycles in order to make the time constant. In both approaches we can see the number of missing bytes are very close; not a very significant difference but in fixed number approach consumes a large amount of CPU clock cycles and therefore the AES implementation becomes slower. Average number approach can be considered as the most fine-tuned version of fixed number approach because in average number approach it is tried to fix the number of clock cycles for encryption to the most optimum value using the average number of clock cycles of encryptions so far.

Finally, we conclude that enforcing constant number of clock cycles is a working countermeasure against cache-timing attacks on AES implementations.